\documentclass[aps,showpacs,preprintnumbers,amsmath,amssymb]{revtex4}

\usepackage{dcolumn}
\usepackage{graphicx}
\usepackage[dvips]{epsfig}
\usepackage{amssymb}
\usepackage{color}
\usepackage{enumerate}
\usepackage{subfigure}
\usepackage{multirow}
\usepackage{diagbox}


\begin{document}
\baselineskip=0.8 cm
%\preprint{example}
\title{\bf The images of Brans-Dicke-Kerr type naked singularities }

\author{Fen Long$^{1}\footnote{Corresponding author: lf@usc.edu.cn}$,
Weike Deng$^{2}$\footnote{wkdeng@hnit.edu.cn},
Xin Qin$^{3,4}$\footnote{qx@hnust.edu.cn},
Songbai Chen$^{5,6}$\footnote{csb3752@hunnu.edu.cn},
Jiliang Jing$^{5,6}$ \footnote{jljing@hunnu.edu.cn}}
\affiliation{ $ ^1$ School of Mathematics and Physics, University of South China,
Hengyang, 421001, People's Republic of China
\\$ ^2$ School of Science, Hunan Institute of Technology, Hengyang 421002, People's Republic of China
\\$ ^3$ School of Physics and Electronic Science, Hunan University of Science and Technology, \
Xiangtan 411021, People's Republic of China
\\$ ^4$ Key Laboratory of Intelligent Sensors and Advanced Sensing Materials of Hunan Province, \
Hunan University of Science and Technology, Xiangtan 411021, People's Republic of China
\\$ ^5$ Department of Physics, Institute of Interdisciplinary Studies, \
\\ Key Laboratory of Low Dimensional Quantum Structures and Quantum Control of Ministry of Education,
\\ Synergetic Innovation Center for Quantum Effects and Applications, Hunan
Normal University,  Changsha, Hunan 410081, People's Republic of China
\\$ ^6$ Center for Gravitation and Cosmology, College of Physical Science and Technology, Yangzhou University, Yangzhou 225009, People's Republic of China}

\begin{abstract}
\baselineskip=0.6 cm
\begin{center}
{\bf Abstract}
\end{center}

We have studied the images of the Brans-Dicke-Kerr spacetime with a dimensionless Brans-Dicke parameter $\omega$, which belongs to axisymmetric rotating solutions in the Brans-Dicke theory. Our results show that the Brans-Dicke-Kerr spacetime with the parameter $\omega>-3/2$ represents naked singularities with distinct structures. For the case with $a \leq M$, the shadow in the Brans-Dicke-Kerr spacetime persists, gradually becomes flatter and smaller as $\omega$ decreases. Especially when $\omega<1/2$, the shadow in the image exhibits a very special ``jellyfish" shape and possesses a self-similar fractal structure. For the case with $a > M$, a distinct gray region consisting of two separate patches appears in the image observed by equatorial observers. This indicating that the Brans-Dicke-Kerr spacetime can be distinguished from the Kerr and Kerr-de Sitter cases based on its image. These effects of the Brans-Dicke parameter could help us to reveal the intrinsic structure of the Brans-Dicke-Kerr spacetimes and provide a foundation for testing Brans-Dicke theory through future high-precision observations.

\end{abstract}

\maketitle
\newpage
\section{Introduction}
The Event Horizon Telescope (EHT) Collaboration has successfully released the images of the supermassive black holes at the centers of the elliptical galaxy M87\cite{EHT1} and the Milky Way Galaxy\cite{EHT2}. This achievement not only marks direct observation of black holes, but also signifies a new era in astrophysics and black hole physics. The formation of the shadow in the image is closely related to the geometric structure and gravitational properties of the spacetime, making it a crucial window for probing strong gravitational fields and testing theories of gravity.

Current observations indicate that Einstein's general relativity successfully accounts for the main features of black hole shadows, while the possibility of modified gravity theories has not been entirely ruled out.
Among various modified theories of gravity, the Brans-Dicke theory is one of the most well-known and extensively studied, which proposes that the Newtonian constant $G$ is not assumed to be a constant but related to the scalar field\cite{BD,BD1}.
The Brans-Dicke theory accommodates Mach's principle\cite{Mach} and Dirac's large-number hypothesis\cite{Dirac}, while successfully reproduces several key predictions of general relativity, such as the precession of planetary perihelia, light deflection, and gravitational redshift\cite{BD}. Consequently, research on the Brans-Dicke theory remains highly active and continues to attract considerable attention.
In the Brans-Dicke theory, the field equations are significantly more intricate than those in general relativity. Therefore, a common strategy is to construct new exact solutions by starting from simpler known solutions in either the Brans-Dicke theory itself or Einstein gravity\cite{BD2,BD3}. Using this solution generating method, various types of solutions in the Brans-Dicke theory have been successfully constructed\cite{BD2,BD3,BD4,BD5,BD6,BD7,BD8}.

Images of compact objects encode valuable information about the spacetime\cite{sw,swo,astro,chaotic,binary,sha18,my,BI,swo7,swo8,swo9,swo10,swo11,swo12,S11,S12,W7,W8,W9,W10,W11,W12,W13,kns,kds,whk1,whk2}, making them a powerful tool for testing various modified theories of gravity\cite{W19,W20,W21,W22,W23,W24,S7,S8,S9,S10}. %检查是否属于修改引力
In this paper, we will study the images of the Brans-Dicke-Kerr spacetime in the Brans-Dicke theory. Hongsu Kim\cite{BD4} derived the Brans-Dicke-Kerr solution directly from explicit spacetime solutions within the Brans-Dicke theory, while Joseph Sultana and Benjamin Bose\cite{BDK} employed the conformal transformation to obtain the corresponding metric in the Einstein frame with a dimensionless Brans-Dicke parameter $\omega$.
In this solution, different values of the parameter $\omega$ correspond to Brans-Dicke-Kerr spacetimes with distinct characteristics\cite{BDK,BDK1}. For certain values of $\omega$, the Brans-Dicke-Kerr metric exhibits finite curvature invariants and zero surface gravity on the horizons $r = r_\pm$, and although the Brans-Dicke scalar field diverges there, it still satisfies the weak energy condition, allowing the metric to describe a nontrivial black hole spacetime distinct from those in Einstein-Maxwell theory\cite{BD4,BDK}. For other values of $\omega$, the Brans-Dicke-Kerr spacetime is characterized by a divergence of the Kretschmann scalar on the horizons $r = r_\pm$, causing the corresponding metric to represent a naked singularity rather than a black hole spacetime\cite{BD4,BDK}.
This class of naked singularity spacetimes characterized by the Kretschmann scalar diverges on the surfaces $r = r_\pm$ also arises in the ``Kerr-like" solution, a reasonable rotating generalization of the Fisher-Janis-Newman-Winicour solution\cite{BG,BG1,BGshadow}. The image of the ``Kerr-like" spacetime with this distinctive structure exhibits a shadow\cite{BGshadow}.
Moreover, recent studies have demonstrated that static spherically symmetric Brans-Dicke naked singularities, despite lacking a conventional event horizon, can still capture photons and produce a shadow smaller than that of a Schwarzschild black hole of the same mass\cite{BDNK}.
Therefore, the image of the Brans-Dicke-Kerr naked singularity with the Brans-Dicke parameter is expected to produce results different from the Kerr case.

The paper is organized as follows. In Sec. II, we briefly review the Brans-Dicke-Kerr spacetime metric and discuss the distinct characteristics of spacetimes exhibiting the secondary curvature singularity, as well as analyze photon motion. In Sec. III, we numerically present the shadows and photon trajectories of the Brans-Dicke-Kerr spacetime and examine the effects of the Brans-Dicke parameter on the shadow. Finally, Sec. IV provides a brief summary of our results.

\section{The Brans-Dicke-Kerr solution}

The Brans-Dicke theory\cite{BD} is a well-known scalar-tensor extension of Einstein's general relativity, in which Newton's gravitational constant is replaced by a scalar field $\varphi$, typically written as $G=1/\varphi$. Its action  is given by the following equation:
\begin{equation}
\begin{aligned}
S=\frac{1}{16 \pi} \int[\varphi R & -\frac{\omega}{\varphi} g^{\alpha \beta} \nabla_\alpha \varphi \nabla_\beta \varphi \left.-V(\varphi)+\mathcal{L}_m\right] \sqrt{-g} d^4 x,
\end{aligned}
\end{equation}
where $\omega$ and $\mathcal{L}_m$ represent the dimensionless Brans-Dicke parameter and the matter action, respectively, while the scalar field potential $V(\varphi)$ is set to zero.
Varying the action with respect to the metric tensor and the scalar field yields the field equations:
\begin{equation}
\begin{aligned}
G_{\mu \nu}  =\frac{8 \pi}{\varphi} T_{\mu \nu}+\frac{\omega}{\varphi^2}\left[\nabla_\mu \varphi \nabla_\nu \varphi-\frac{1}{2} g_{\mu \nu} \nabla_\alpha \varphi \nabla^\alpha \varphi\right] +\frac{1}{\varphi}\left[\nabla_\mu \nabla_\nu \varphi-g_{\mu \nu} \square \varphi\right],
\end{aligned}
\end{equation}
\begin{equation}
\square \varphi=\frac{8 \pi}{2 \omega+3} T^{(m)},\label{field equations2}
\end{equation}
where $T_{\mu \nu}=\frac{-2}{\sqrt{-g}} \frac{\delta}{\delta g^{\mu \nu}}\left(\sqrt{-g} \mathcal{L}_m\right)$ and $T^{(m)}=T^{\mu}_{\mu}$ represent the energy-momentum tensor of matter and its trace, respectively.

By solving the above field equations, a Kerr-like solution in Brans-Dicke theory under the Jordan frame can be obtained\cite{BDK,BD4}, with the scalar field expressed as $\varphi=(\Delta/M^{2})^{2 /(2 \omega+3)} \sin ^{4 /(2 \omega+3)} \theta $. Through the conformal transformation $g_{a b} \rightarrow \tilde{g}_{a b}=\Omega^2 g_{a b}$, with $\Omega=\sqrt{G\varphi}$, the solution can be expressed in the Einstein frame, and the scalar field is then redefined as
\begin{equation}
\tilde{\varphi}=\sqrt{\frac{2 \omega+3}{16 \pi G}} \ln \left(\frac{\varphi}{\varphi_0}\right),\label{scalar field}
\end{equation}
where $\varphi_0$ is the current value of the gravitational constant.
Substituting the scalar field from the Jordan frame into its definition in the Einstein frame  Eq.(\ref{scalar field}) yields:
\begin{equation}
\tilde{\varphi}=\frac{1}{2 \sqrt{\pi} \sqrt{2 \omega+3}} \ln \left(\frac{\Delta}{M^2} \sin ^2 \theta\right).
\end{equation}
To ensure that the transformed scalar field $\tilde{\varphi}$ remains real, we consider only $\omega>-3/2$ in the following discussion. The corresponding field equation takes a simple form
\begin{equation}
\begin{aligned}
R_{\mu \nu} & =\frac{1}{2} \nabla_\mu \tilde{\varphi} \nabla_\nu \tilde{\varphi}, \\
\square \tilde{\varphi} & =0.
\end{aligned}
\end{equation}
Using the metric in the Jordan frame along with the conformal transformation, one obtains the metric in the Einstein frame\cite{BDK}:
\begin{equation}
\begin{aligned}
d s^2=& -\left( 1-\frac{2 M r}{\Sigma} \right) d t^2-\frac{4 M a r \sin ^2 \theta }{\Sigma} d t d \psi+\left(r^2+a^2+\frac{2 M a^2 r}{\Sigma} \sin ^2 \theta\right) \sin ^2 \theta d \psi^2  \\
& +\left(\frac{\Delta \sin ^2 \theta}{M^2} \right)^{4/\left(2\omega+3\right)} \Sigma\left(\frac{d r^2}{\Delta}+d \theta^2\right),\label{BDK}
\end{aligned}
\end{equation}
with
\begin{equation}
\begin{aligned}
\Sigma=r^2+a^2 \cos ^2 \theta, \quad  \Delta & =r^2+a^2-2 M r,
\end{aligned}
\end{equation}
where $M$ and $a$ represent the mass and the rotation parameter, respectively. As the Brans-Dicke parameter $\omega \rightarrow \infty$, this axisymmetric rotating solution in Brans-Dicke theory reduces to the Kerr solution. The presence of the Brans-Dicke parameter $\omega$ further influences the image by modifying the spacetime properties throughout the entire region.

Next, we introduce the two key surfaces in the strong gravity region of the Brans-Dicke-Kerr solution that significantly influence the formation of images: the horizon and the curvature singularity. For Brans-Dicke parameter $\omega>1/2$, the horizon radii $r_{\pm}$ of the Brans-Dicke-Kerr solution are given by the roots of the following equation
\begin{eqnarray}
\Delta = r^2+a^2-2 M r& =& 0.
\end{eqnarray}
The Kretschmann scalar $\kappa$ to determine the location of the curvature singularity, denoted by
\begin{equation}
\kappa=R_{\mu \nu \rho \sigma} R^{\mu \nu \rho \sigma}=\Sigma^{-6}\Delta^{-\left(4\omega+14\right)/\left(2\omega+3\right)} g(r, \theta, a, M, \omega) .\label{kappa}
\end{equation}
As the expression $g(r, \theta, a, M, \omega)$ is complicated, we do not display it here.
The Kretschmann scalar indicates that the Brans-Dicke-Kerr solution possesses the same intrinsic curvature singularity as the Kerr solution at $\Sigma=0$, which lies on the ring at $r=0,\theta=\pi/2$. Moreover, it also indicates that the Brans-Dicke-Kerr solution possesses a secondary curvature singularity at $\Delta=0$ when the parameters satisfy $a\leq M$ and $-\left(4\omega+14\right)/\left(2\omega+3\right)<0$.
Therefore, When the rotation parameter $a\leq M$, two distinct cases arise in the spacetime depending on the value of $\omega$. For the Brans-Dicke parameter $\omega>1/2$, the divergence of the Kretschmann scalar $\kappa$ at $\Delta=0$ leads to a secondary curvature singularity at the horizon radii $r_{\pm}$, so that the Brans-Dicke-Kerr metric describes a naked singularity. For the Brans-Dicke parameter $-3/2<\omega\leq 1/2$, the disappearance of the horizon causes spacetime to become a naked singularity with the ring singularity located at $\Sigma=0$ and the secondary curvature singularity at $\Delta=0$.
When the rotation parameter $a>M$, $\Delta=0$ has no real roots, resulting in the absence of both the horizon and the secondary curvature singularity, so the spacetime corresponds to a naked singularity containing only a ring singularity for any Brans-Dicke parameter $\omega$.
To conclude, in the Brans-Dicke-Kerr spacetime, naked singularities can occur for both the rotation parameter $a\leq M$ and $a>M$, whereas in the Kerr spacetime they occur only when $a>M$. Moreover, for the rotation parameter $a>M$, the naked singularity in the Brans-Dicke-Kerr spacetime behaves like a Kerr naked singularity with hair $\omega$. Since the formation mechanisms of the three types of naked singularities in the Brans-Dicke-Kerr spacetime differ from that of the Kerr naked singularity, their images are expected to display distinct features.

We now examine the motion of photons in the Brans-Dicke-Kerr spacetime (\ref{BDK}). In a curved spacetime, the Hamiltonian for photons propagating along null geodesics can be expressed as
\begin{eqnarray}
 H(x,p)=\frac{1}{2}g^{\mu \nu}(x)p_{\mu}p_{\nu}=0.\label{hamiltonian}
\end{eqnarray}
There are two conserved quantities: the energy $E$ and the angular momentum $L_z$ with the following expressions
\begin{eqnarray}
E=-p_{t}=-g_{tt}\dot{t}-g_{t\psi}\dot{\psi}, \quad L_{z}=p_{\phi}=g_{t\psi}\dot{t}+g_{\psi\psi}\dot{\psi}.\label{conserved quantities}
\end{eqnarray}
From these conserved quantities we obtain the photon equations of motion along null geodesics
\begin{eqnarray}
\dot{t}&=&\frac{g_{\psi\psi}E+g_{t\psi}L_z}{g_{t\psi}^2-g_{tt}g_{\psi\psi}},\label{u1}\\
\dot{\psi}&=&\frac{g_{t\psi}E+g_{tt}L_z}{g_{tt}g_{\psi\psi}-g_{t\psi}^2},\label{u4}\\
\ddot{r}&=&\frac{1}{2g_{rr}}(g_{tt,r}\dot{t}^2-g_{rr,r}\dot{r}^2+g_{\theta\theta,r}\dot{\theta}^2+g_{\phi\phi,r}\dot{\phi}^2
+2g_{t\psi,r}\dot{t}\dot{\psi}-2g_{\theta\theta,\theta}\dot{r}\dot{\theta}),\label{uu2}\\
\ddot{\theta}&=&\frac{1}{2g_{\theta\theta}}(g_{tt,\theta}\dot{t}^2+g_{rr,\theta}\dot{r}^2-g_{\theta\theta,
\theta}\dot{\theta}^2+g_{\psi\psi,\theta}\dot{\psi}^2+2g_{t\psi,\theta}\dot{t}\dot{\psi}-2g_{\theta\theta,r}
\dot{r}\dot{\theta}).\label{uu3}
\end{eqnarray}
From Eqs.(\ref{uu2}) and (\ref{uu3}), it can be seen that $\ddot{r}$ and $\ddot{\theta}$ both depend on Brans-Dicke parameter $\omega$, because the metric components $g_{rr}$ and $g_{\theta\theta}$ contain $\omega$. This causes photon trajectories in the Brans-Dicke-Kerr spacetime to deviate from those in the Kerr spacetime, thereby affecting the images.

\section{The images of the Brans-Dicke-Kerr spacetime}
In the following, we investigate image formation in the Bran-Dicke-Kerr and examine the influence of the Brans-Dicke parameter $\omega$ on the images.
We employ the ``backward ray-tracing" method\cite{sw,swo,astro,chaotic,binary,sha18,my,BI,swo7,swo8,swo9,swo10} to numerically simulate the shadow of the Brans-Dicke-Kerr spacetime. In this approach, light rays are traced backward from the observer by numerically integrating the null geodesic equations(\ref{u1})-(\ref{uu3}), which determines the position of each pixel in the final image. The observer's basis $\{e_{\hat{t}},e_{\hat{r}},e_{\hat{\theta}},e_{\hat{\psi}}\}$ can be expressed in terms of the coordinate basis$\{\partial_t,\partial_r,\partial_{\theta},\partial_{\psi} \}$.
\begin{eqnarray}
\label{zbbh}
e_{\hat{\mu}}=e^{\nu}_{\hat{\mu}} \partial_{\nu},
\end{eqnarray}
where the matrix $e^{\nu}_{\hat{\mu}}$ satisfies $g_{\mu\nu}e^{\mu}_{\hat{\alpha}}e^{\nu}_{\hat{\beta}}
=\eta_{\hat{\alpha}\hat{\beta}}$, and $\eta_{\hat{\alpha}\hat{\beta}}$ denotes the usual Minkowski metric. For the Brans-Dicke-Kerr spacetime (\ref{BDK}), it is convenient to choose a decomposition \cite{sw,swo,astro,chaotic,binary,sha18,my,BI,swo7,swo8,swo9,swo10,swo11,swo12}
\begin{eqnarray}
\label{zbbh1}
e^{\nu}_{\hat{\mu}}=\left(\begin{array}{cccc}
\zeta&0&0&\gamma\\
0&A^r&0&0\\
0&0&A^{\theta}&0\\
0&0&0&A^{\psi}
\end{array}\right),
\end{eqnarray}
where $\zeta$, $\gamma$, $A^r$, $A^{\theta}$,and $A^{\phi}$ are real coefficients.
From the Minkowski normalization condition
\begin{eqnarray}
e_{\hat{\mu}}e^{\hat{\nu}}=\delta_{\hat{\mu}}^{\hat{\nu}},
\end{eqnarray}
it follows that
\begin{eqnarray}
\label{xs}
&&A^r=\frac{1}{\sqrt{g_{rr}}},\;\;\;\;\;\;\;
A^{\theta}=\frac{1}{\sqrt{g_{\theta\theta}}},\;\;\;\;\;\;\;
A^{\psi}=\frac{1}{\sqrt{g_{\psi\psi}}},\;\;\;\;\;\;\;\nonumber\\
&&\zeta=\sqrt{\frac{g_{\psi\psi}}{g_{t\psi}^{2}-g_{tt}g_{\psi\psi}}},\;\;\;\;\;\;
\gamma=-\frac{g_{t\psi}}{g_{\psi\psi}}\sqrt{\frac{g_{\psi\psi}}{g_{t\psi}^{2}-g_{tt}g_{\psi\psi}}}.
\end{eqnarray}
According to Eq.(\ref{zbbh}), the locally measured four-momentum $p^{\hat{\mu}}$ of a photon takes the form
\begin{eqnarray}
\label{dl}
p^{\hat{t}}=-p_{\hat{t}}=-e^{\nu}_{\hat{t}} p_{\nu},\;\;\;\;\;\;\;\;\;
\;\;\;\;\;\;\;\;\;\;\;p^{\hat{i}}=p_{\hat{i}}=e^{\nu}_{\hat{i}} p_{\nu}.
\end{eqnarray}
By making use of Eq.(\ref{xs}), the locally measured four-momentum $p^{\hat{\mu}}$ in the Brans-Dicke-Kerr spacetime can be derived as follows
\begin{eqnarray}
\label{kmbh}
p^{\hat{t}}&=&\zeta E-\gamma L,\;\;\;\;\;\;\;\;\;\;\;\;\;\;\;\;\;\;\;\;
p^{\hat{r}}=\frac{1}{\sqrt{g_{rr}}}p_{r},\nonumber\\
p^{\hat{\theta}}&=&\frac{1}{\sqrt{g_{\theta\theta}}}p_{\theta},
\;\;\;\;\;\;\;\;\;\;\;\;\;\;\;\;\;\;\;\;\;\;
p^{\hat{\psi}}=\frac{1}{\sqrt{g_{\psi\psi}}}L,
\end{eqnarray}
Hence, the celestial coordinates corresponding to a given light ray in the spacetime (\ref{BDK}) are written as
\begin{eqnarray}
\label{xd1}
x&=&-r_{obs}\frac{p^{\hat{\psi}}}{p^{\hat{r}}}
=-r_{obs}\sqrt{\frac{1}{g_{rr}g_{\psi\psi}}}\frac{g_{t\psi}\dot{t}+g_{\psi\psi}\dot{\psi}}{\dot{r}}, \nonumber\\
y&=&r_{obs}\frac{p^{\hat{\theta}}}{p^{\hat{r}}}=
r_{obs}\sqrt{\frac{g_{\theta\theta}}{g_{rr}}}\frac{\dot{\theta}}{\dot{r}},
\end{eqnarray}
where $r_{obs}, \theta_{obs}$ are the radial coordinate and polar angle of observer.

\begin{figure}[t]
\begin{center}
\includegraphics[width=4cm]{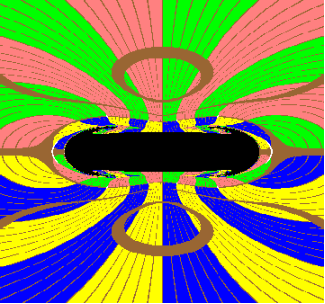}\includegraphics[width=4cm]{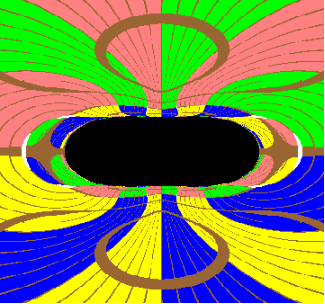}\includegraphics[width=4cm]{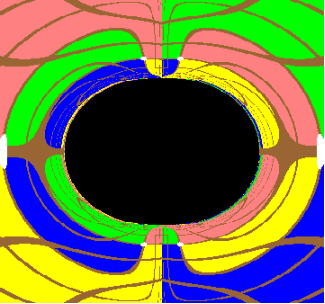}\includegraphics[width=4cm]{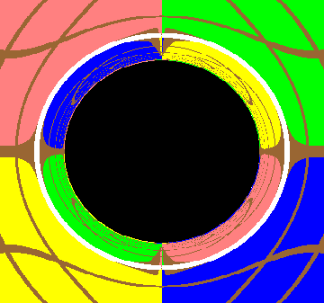}
\caption{ The variation of images with the Brans-Dicke parameter $\omega$ for the Brans-Dicke-Kerr spacetime with fixed $a=0$. Here we set the mass parameter $M=1$, $r_{obs}=8M$ and $\theta_{obs}=\pi/2$. The figures from left to right correspond to $\omega=0.4$, $1$, $10$, and $500$, respectively.}\label{Figa0}
\end{center}
\end{figure}

\begin{figure}[t]
\begin{center}
\includegraphics[width=4cm]{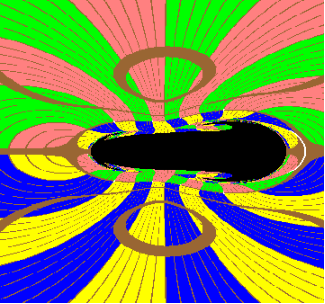}\includegraphics[width=4cm]{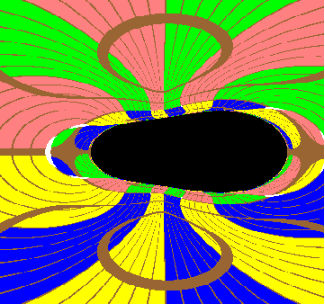}\includegraphics[width=4cm]{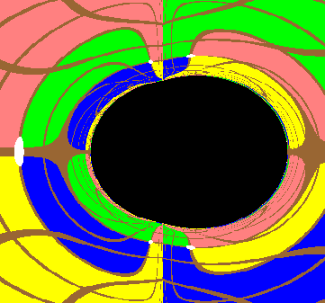}\includegraphics[width=4cm]{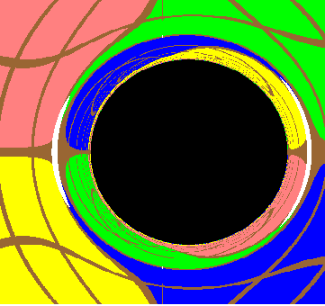}
\caption{ The variation of images with the Brans-Dicke parameter $\omega$ for the Brans-Dicke-Kerr spacetime with fixed $a=0.5$. Here we set the mass parameter $M=1$, $r_{obs}=8M$ and $\theta_{obs}=\pi/2$. The figures from left to right correspond to $\omega=0.4$, $1$, $10$, and $500$, respectively.}\label{Figa05}
\end{center}
\end{figure}

\begin{figure}[t]
\begin{center}
\includegraphics[width=4cm]{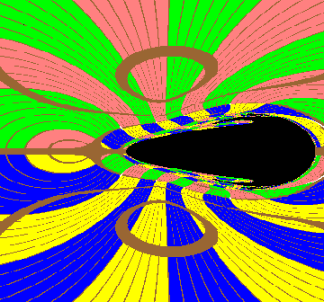}\includegraphics[width=4cm]{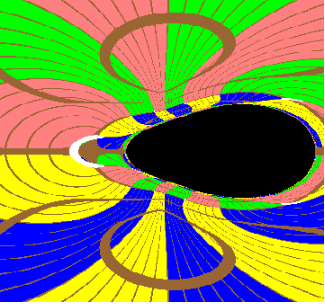}\includegraphics[width=4cm]{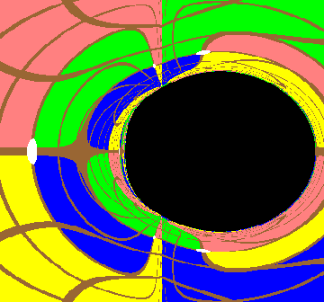}\includegraphics[width=4cm]{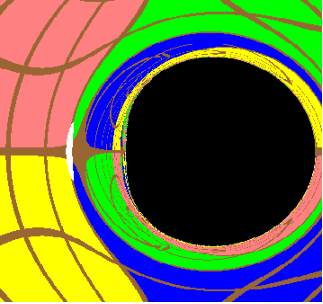}
\caption{ The variation of shadows with the Brans-Dicke parameter $\omega$ for the Brans-Dicke-Kerr-type black holes and naked singularities with fixed $a=0.99$. Here we set the mass parameter $M=1$, $r_{obs}=8M$ and $\theta_{obs}=\pi/2$. The figures from left to right correspond to $\omega=0.4$, $1$, $10$, and $500$, respectively.}\label{Figa099}
\end{center}
\end{figure}

\begin{figure}[t]
\begin{center}
\includegraphics[width=4cm]{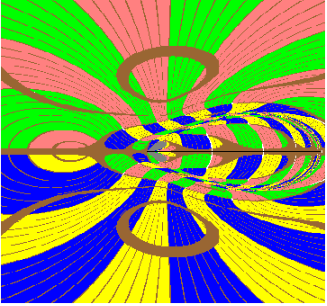}\includegraphics[width=4cm]{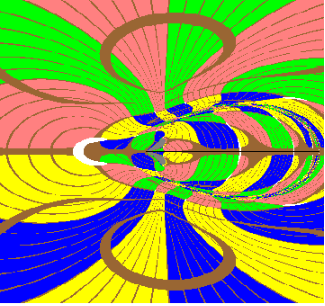}\includegraphics[width=4cm]{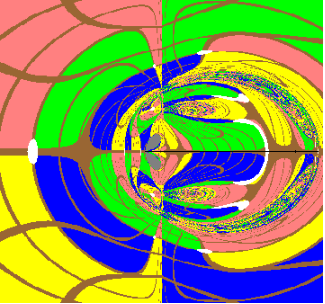}\includegraphics[width=4cm]{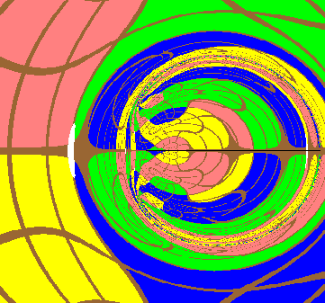}
\caption{ The variation of images with the Brans-Dicke parameter $\omega$ for the Brans-Dicke-Kerr spacetime with fixed $a=1.05$. Here we set the mass parameter $M=1$, $r_{obs}=8M$ and $\theta_{obs}=\pi/2$. The figures from left to right correspond to $\omega=0.4$, $1$, $10$, and $500$, respectively.}\label{Figa105}
\end{center}
\end{figure}

Figs.\ref{Figa0}-\ref{Figa105} present the images of the Brans-Dicke-Kerr spacetime obtained by an observer located on the equatorial plane for different values of the rotation parameter $a$ and the Brans-Dicke parameter $\omega$.
For the rotation parameter $a\leq M$, the shadow in the Brans-Dicke-Kerr spacetime always exists and enlarges with increasing $\omega$.
As shown in Fig.\ref{Figa0} for the rotating spacetime cases with $a=0$, when the Brans-Dicke-Kerr parameter $\omega>1/2$ decreases (where the event horizon coincides with the secondary curvature singularity), the shadow in the Brans-Dicke-Kerr spacetime gradually deviates from the perfect disk shape of the Kerr black hole, becoming a flattened elliptical disk while still maintaining left-right symmetry.
When the Brans-Dicke-Kerr parameter $-3/2<\omega\leq 1/2$ (where the event horizon disappears but the secondary curvature singularity still exists), the shadow will exhibit a very special ``two-headed jellyfish" shape.
As shown in Figs.\ref{Figa05}-\ref{Figa099} for the rotating spacetime cases with $a=0.5$ and $a=0.99$, as the Brans-Dicke-Kerr parameter $\omega$ decreases the shadow gradually flattens and shrinks while the left half showing more obvious changes than the right half. The overall shape of shadow gradually takes on a footprint-like shape and eventually develops into a ``jellyfish" shape.
Therefore, for $a\leq M$, the Brans-Dicke-Kerr solutions belong to a class of spacetimes that exhibit naked singularities. Although these solutions lack a conventional event horizon, they still give rise to shadow-like features. Similar phenomena also occur in the spherically symmetric metric belong to the Brans-Dicke gravity\cite{BDNK} and the Bogush-Galt'sov naked singularity\cite{BGshadow}. These results suggest that there remains the possibility for naked singularities to exist as compact astrophysical objects.
For the rotation parameter $a>M$, the naked singularity in the Brans-Dicke-Kerr spacetime behaves like a Kerr naked singularity with hair $\omega$.
As shown in Fig.\ref{Figa105}, similar to the Kerr naked singularity\cite{kns}, the Brans-Dicke-Kerr naked singularity lacks an event horizon and has only a ring singularity, so the image shows a bright background light source with a black straight line rather than a closed black shadow region.
Unlike in the case of a Kerr naked singularity, as the Brans-Dicke parameter $\omega$ decreases, a distinct gray region appears near the center on the left side of the Brans-Dicke-Kerr spacetime naked singularity image obtained by equatorial observers.
In this region, photons originating from the negative radii region can traverse $r=0, \theta \neq \pi /2$ and reach observers in the positive radii region.
If we assume that a sphere light source exists in the negative radii region of the Brans-Dicke-Kerr naked singularity spacetime, the gray region corresponds to a bright region illuminated by photons originating from the negative radii region. Conversely, without a light source, the gray region appears as the completely dark spot.
The presence of gray regions depends not only on the rotation parameters but also on the inclination angle of the observer. For observers deviates from the equatorial plane, dark spot appear in the Kerr naked singularity image, whereas in the Brans-Dicke-Kerr naked singularity spacetime they will shrink.
The study of the Kerr-de Sitter naked singularities also takes into account photons entering the negative radii region of the spacetime\cite{kds}. As a result the shadow of the Kerr-de Sitter naked singularities on the observer's celestial sky includes not only an arc or circle structure but also a dark spot corresponding to photons escaping to negative infinity\cite{kds}. Since the gray regions in the image of the Brans-Dicke-Kerr naked singularity appear as two separate patches rather than as a single dark spot as in the Kerr naked singularities\cite{kns} and the Kerr-de Sitter naked singularities\cite{kds}, it indicates that the Brans-Dicke-Kerr naked singularity can be distinguished from the Kerr and the Kerr-de Sitte cases based on its image.
To clarify the formation of the shadow in the Brans-Dicke-Kerr spacetime image, we further investigate the trajectory of photon motion.

\begin{figure}[t]
\begin{center}
\subfigure[$\omega=10$]{\includegraphics[width=6cm]{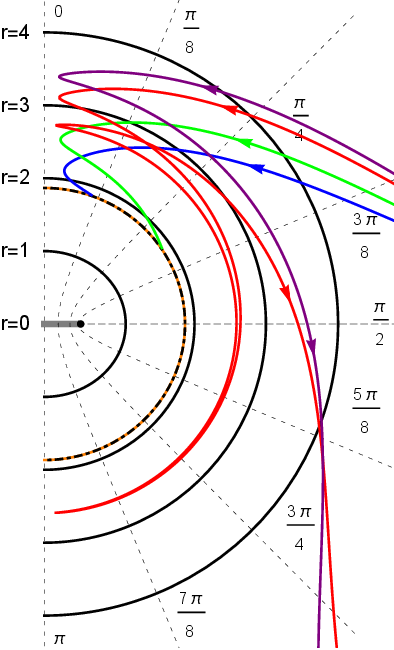}}\;\;\;\;\;\;\subfigure[$\omega=0.4$]{\includegraphics[width=6cm]{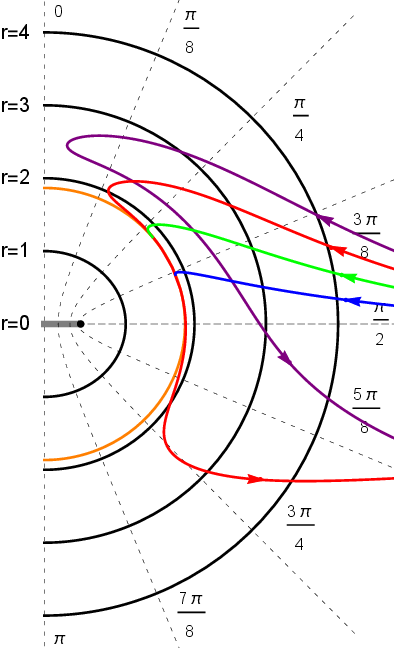}}
\caption{Photon trajectories in the $(r-\theta)$ plane for the Brans-Dicke-Kerr spacetime with $a=0.5$. The black half-ellipses denote surfaces of constant Boyer-Lindquist radius $r$, while the black dashed hyperbolas correspond to surfaces of constant latitude coordinate $\theta$.
The black dot-dashed line represents the event horizon, the orange solid line indicates the secondary curvature singularity, and the other colored solid curves show photon trajectories for $x=-0.3$ with different values of $y$. In the left panel, the purple, red, green, and blue curves are associated with $y=4.5, 4.168, 3.5, 3.0$; in the right panel, they take values $y=1.6, 1.2, 0.8, 0.4$.}\label{Figrt1}
\end{center}
\end{figure}

\begin{figure}[t]
\begin{center}
{\includegraphics[width=6cm]{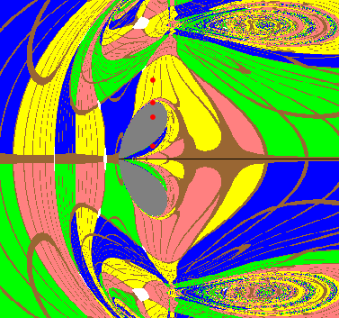}}\;\;\;\;\;\;{\includegraphics[width=5.5cm]{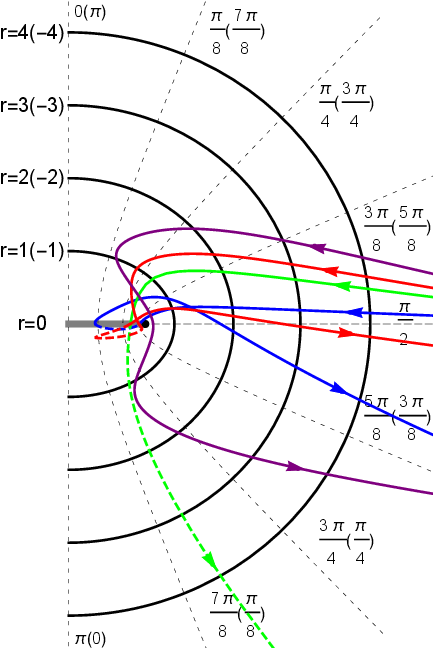}}
\caption{Photon trajectories in the $(r-\theta)$ plane for the Brans-Dicke-Kerr spacetime with $a=1.05,\omega=10$ under different initial conditions. Black half-ellipses denote surfaces of constant Boyer-Lindquist radius $r$(with $r<0$ cases indicated in parentheses), while black dashed hyperbolas correspond to surfaces of constant latitude coordinate $\theta$(with the distribution of these surfaces for $r<0$ shown in parentheses). The colored curves represent photon trajectories for $x=-0.3$ with different values of $\beta$(purple, red, green, and blue correspond to $y=1.5, 1.074, 0.8, 0.25$), corresponding to the four red points marked on the image of the left panel. The solid portions of the curves correspond to radius $r>0$, while dashed portions correspond to $r<0$.}\label{Figrt2}
\end{center}
\end{figure}

Figs.\ref{Figrt1}-\ref{Figrt2} present photon trajectories in the $(r-\theta)$ plane with different Brans-Dicke parameter and rotation parameter in the Brans-Dicke-Kerr spacetime. The projection of the photon trajectories onto the $(r-\theta)$ plane is obtained from the transformation between Cartesian and Boyer-Lindquist coordinates, with the horizontal coordinate expressed as $X=\sqrt{r^2+a^2}\sin{\theta}$ and the vertical coordinate as $Z=r\cos{\theta}$.
In Figs.\ref{Figrt1}(a-b), we find that in the Brans-Dicke-Kerr spacetime with a rotation parameter of $a=0.5$, photons can be captured near the naked singularity whether the event horizon coincides with the secondary curvature singularity or the event horizon disappears while the secondary curvature singularity exists alone. As a result, shadows appear in the image for both configurations of the naked singularity.
As the Brans-Dicke parameter $\omega$ decreases, the minimum turning points in the radial direction gradually shrink and approach the secondary singularity, which implies smaller shadows.
In Figs.\ref{Figrt2}, we find that in the Brans-Dicke-Kerr spacetime with a rotation parameter of $a=1.05$, where both the event horizon and the secondary curvature singularity are absent.
In the ``backward ray-tracing" method, part of photons are traced from the observer and pass through the positive radii region, approaching the radial coordinate $r=0$ but reaching a latitudinal coordinate $\theta \neq \pi/2$, thereby bypassing the ring singularity and entering the negative radii region. Photons with a radial turning point return to the positive radii region, while photons without such turning point escape to negative infinity, resulting in gray region appearing in the image.
Due to the crucial role of photon motion in forming spacetime images, the photon dynamics in Brans-Dicke-Kerr naked singularities differ from those in Kerr naked singularities and generate distinct gray regions in the image observed by an equatorial-plane observer.

\begin{figure}[t]
\begin{center}
{\includegraphics[width=5.4cm]{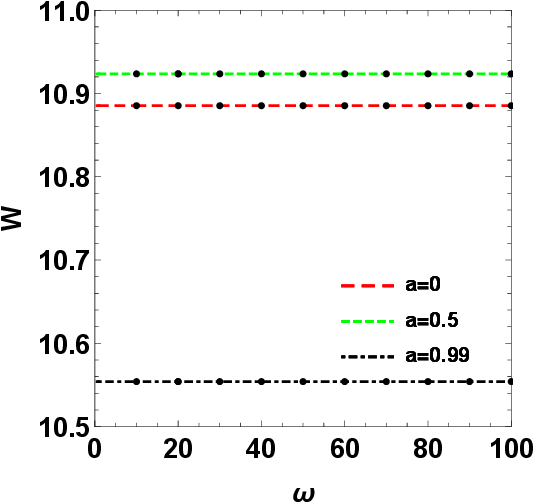}}\;{\includegraphics[width=5.2cm]{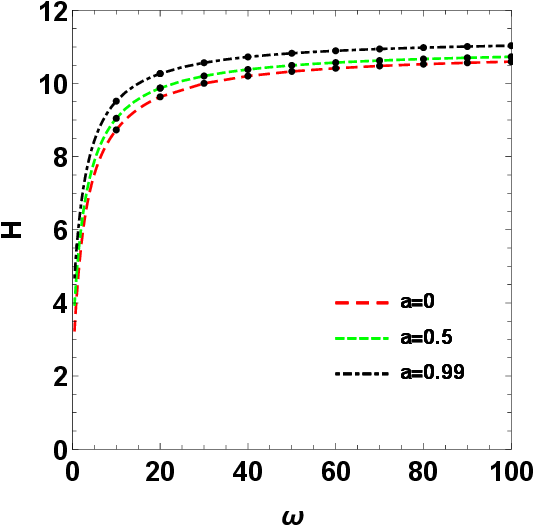}}
\;{\includegraphics[width=5.3cm]{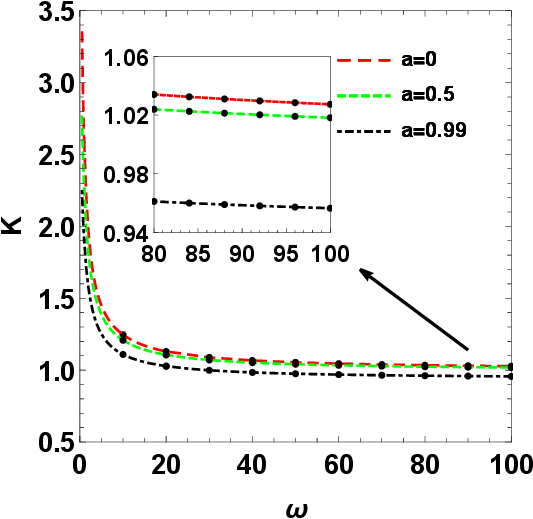}}
\caption{The change in the shadow width $W$, the height $H$ and the oblateness $K$ with the Brans-Dicke parameter $\omega$ in the Brans-Dicke-Kerr spacetime for different rotation parameter $a$.}\label{WHK}
\end{center}
\end{figure}
In Figs.\ref{WHK}, we shows the variations of the width $W$, the height $H$ and the oblateness $K$ of the shadow with respect to the Brans-Dicke parameter $\omega$ for different values of the rotation parameter $a$ in the Brans-Dicke-Kerr spacetime. The quantities $W$, $H$ and $K$ are defined as\cite{whk1,whk2}
\begin{eqnarray}
W=x_r-x_l,\quad\quad\quad\quad H=y_t-y_b, \quad\quad\quad\quad K=\frac{W}{H},
\end{eqnarray}
where $y_t$ and $y_b$ represents the vertical coordinates of the topmost and the bottommost point of the shadow, while $(x_r, 0)$ and $(x_l,0)$ denote its intersections with the horizontal axis on the right and left sides, respectively.

Figs.\ref{WHK} shows that for a fixed rotation parameter $a$, as the Brans-Dicke parameter $\omega$ decreases, the shadow width $W$ remains unchanged as in the Kerr case. Meanwhile, the height $H$ gradually decreases, causing the shadow to appear more flattened and thus increasing its oblateness $K$, which is consistent with the behavior observed in the image. For a fixed Brans-Dicke parameter $\omega$, as the rotation parameter $a$ increases, the shadow width $W$ first increases and then decreases, while the height $H$ increases and the oblateness $K$ gradually decreases. It is observed that the shadow shape varies more significantly with the Brans-Dicke parameter $\omega$ than with the rotation parameter $a$.

\begin{figure}[t]
\begin{center}
\includegraphics[width=4.3cm]{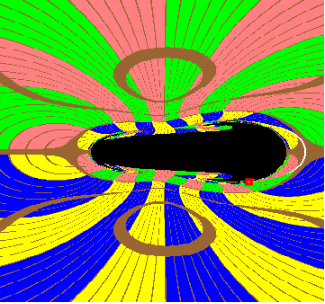}\includegraphics[width=4cm]{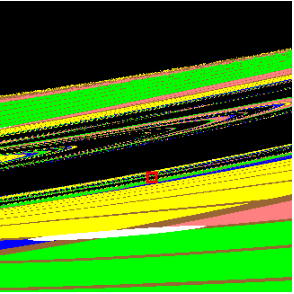}\includegraphics[width=4cm]{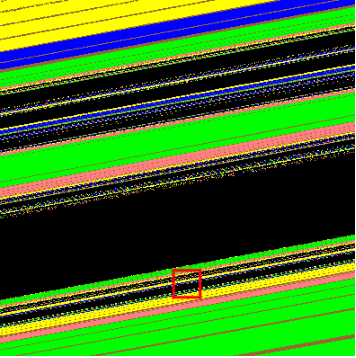}\includegraphics[width=4cm]{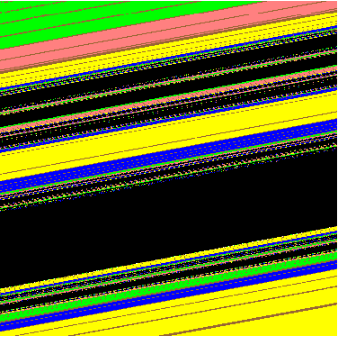}
\caption{The two-headed jellyfish-shaped shadow and the self-similar fractal structures in the shadow of the Brans-Dicke-Kerr spacetime with fixed $a=0.5$ and $\omega=0.4$. Here we set the mass parameter $M=1$, $r_{obs}=8M$ and $\theta_{obs}=\pi/2$.}\label{Figa05chaos}
\end{center}
\end{figure}

As shown in Figs.\ref{Figa0}-\ref{Figa099}, when $a\leq M$ and $\omega<1/2$, the shadow of the Brans-Dicke-Kerr naked singularity exhibits a `jellyfish" shape that is symmetrically distributed about the horizontal line. Within this `jellyfish" shape shadow, many smaller eyebrow-like features can also be observed as shown in Figs.\ref{Figa05chaos}. This indicates that the shadow of the axisymmetric Brans-Dicke-Kerr naked singularity possesses a self-similar fractal structure caused by chaotic photon scattering, a property that is different from that of the Kerr spacetime. These new shadow features arising from the Brans-Dicke parameter help us understand the Brans-Dicke-Kerr naked singularities and provide a theoretical basis for testing the nature of gravity in the future.

\section{Summary}
In this paper we have investigated the images of Brans-Dicke-Kerr type naked singularities with the Brans-Dicke-Kerr parameter $\omega$, analyzed the influence of parameter $\omega$ on the spacetime structure, the shadows, and the photon trajectories.

Firstly, We analyzed the distribution of the horizon and the secondary curvature singularity, which are important for the formation of shadows. When the rotation parameter satisfies $a \leq M$, the Brans-Dicke parameter plays a crucial role in determining the spacetime structure. For $\omega > 1/2$, the horizon coincides with the secondary curvature singularity at $r_\pm$, and the metric describes a naked singularity rather than a black hole. When $-3/2 < \omega \leq 1/2$, the horizon disappears while the secondary curvature singularity remains, so the Brans-Dicke-Kerr spacetime still represents a naked singularity. When the rotation parameter $a > M$, both the horizon and the secondary curvature singularity vanish, and the Brans-Dicke-Kerr naked singularity behaves like a Kerr naked singularity with the hair $\omega$. Since the formation of the three types of Brans-Dicke-Kerr naked singularities differs from that of the Kerr case, their images are expected to exhibit distinct features.

Our result show that for the rotation parameter $a\leq M$, the shadow in the Brans-Dicke-Kerr spacetime always exists. As the Brans-Dicke-Kerr parameter $\omega$ decreases, the shadow gradually becomes flatter and smaller, eventually evolving into a ``jellyfish" shape. Although the Brans-Dicke-Kerr solutions lack a conventional event horizon, they still produce shadow-like features. Similar phenomena also occur in the spherically symmetric metric belong to the Brans-Dicke gravity\cite{BDNK} and the Bogush-Galt'sov naked singularity\cite{BGshadow}. These results suggest that naked singularities may still exist as compact astrophysical objects. For the rotation parameter $a > M$, a distinct gray region appears near the center of the image observed by equatorial observers. Photons traced backward from the observer approach $r=0$ at $\theta \neq \pi/2$ and bypass the ring singularity to escape to negative infinity, forming a gray region in the image. Unlike the single dark spot in Kerr\cite{kns} and Kerr-de Sitter naked singularities\cite{kds}, the Brans-Dicke-Kerr case shows two separate gray patches, indicating that it can be distinguished from the Kerr and Kerr-de Sitter cases based on its image.

Finally, we calculated the shadow's width $W$, height $H$, and oblateness $K$. For a fixed rotation parameter $a$, as the Brans-Dicke parameter $\omega$ decreases, the width $W$ remains unchanged, while the height $H$ gradually decreases, leading to an increase in oblateness $K$.
For a fixed $\omega$, as the rotation parameter $a$ increases, the width $W$ first increases and then decreases, whereas the height $H$ increases and the oblateness $K$ gradually decreases. When $a\leq M$ and $\omega<1/2$, the shadow of the Brans-Dicke-Kerr naked singularity possesses a self-similar fractal structure caused by chaotic photon scattering, a property that is essentially different from that of the Kerr spacetime.

In conclusion, the Brans-Dicke-Kerr spacetime with the Brans-Dicke parameter $\omega>-3/2$ represents different types of naked singularities. For the rotation parameter $a \leq M$, the shadow in the Brans-Dicke-Kerr spacetime persists and gradually becomes flatter and smaller as $\omega$ decreases. For $a > M$, a distinct gray region appears in the image observed by equatorial observers. Especially, for $a\leq M$ and $\omega<1/2$, the shadow of the Brans-Dicke-Kerr naked singularity exhibit a very special ``jellyfish" shape and possesses a self-similar fractal structure. Such investigations not only help reveal the intrinsic structure of the Brans-Dicke-Kerr spacetimes but also provide essential theoretical foundations for testing Brans-Dicke theory through future high-precision observations.

\section{\bf Acknowledgments}
This work was partially supported by the National Natural
Science Foundation of China (Grant Nos. 12205140, 12405053, 12275078, 11875026, 12035005, and 2020YFC2201400)
and the Natural Science Foundation of Hunan Province (Grant No. 2023JJ40523 and 2024JJ6211).

\end{document}